\documentstyle[12pt]{article} 
\begin{document} 
\def \be {\begin{equation}} 
\def \ee {\end{equation}}

\begin{titlepage}

\title{Static chaos and scaling behaviour in the spin-glass phase}
\author{Felix Ritort}

\maketitle

\begin{center}
{\small\baselineskip=14pt Dipartimento di Fisica, Universit\`a di Roma
II, ''Tor Vergata'',\\Viale della Ricerca Scientifica, 00133 Roma, Italy}
\end{center}

\vskip 10mm
\quad     Short title: Static chaos in spin glasses
\vskip 10mm

\par
\quad PACS. 75.5 0 L-- Spin glasses.
\vskip .1in
\begin{abstract} 
We discuss the problem of static chaos in spin glasses. In the case of
magnetic field perturbations, we propose a scaling theory for the
spin-glass phase. Using the mean-field approach we argue that some pure
states are suppressed by the magnetic field and their free energy cost
is determined by the finite-temperature fixed point exponents. In this
framework, numerical results suggest that mean-field chaos exponents are
probably exact in finite dimensions.
If we use the droplet approach, numerical results suggest that the
zero-temperature fixed point exponent $\theta$ is very close to
$\frac{d-3}{2}$. In both approaches $d=3$ is the lower critical
dimension in agreement with recent numerical simulations.
\end{abstract}
\begin{flushright}
{\bf cond-mat/9404020}
\end{flushright}

\end{titlepage}

\baselineskip 6mm

\newpage 

\section{Introduction}

One of the most interesting open problems in spin glasses regards a
correct understanding of the nature of the low temperature phase, i.e.
the spin-glass phase \cite{Sp,Bi86}. Spin glasses are characterized by a
strong freezing at a certain critical temperature.  Below that
temperature a complete description of the nature of the static phase is
still missing.

During the last years there have been several developments in the field,
the most well known being the mean-field theory \cite{Sk75}. Unfortunately
mean-field theory has revealed a complex theoretical structure which is
very obscure when applied to non exactly solvable models for
which some kind of perturbation theory is needed.

There are other approaches to spin-glasses which are known as
phenomenological droplet models, a complete description of them has been
given by D. S Fisher and D. Huse \cite{Fi88,Ko88}. The main idea
underlying these models is that the spin-glass behaviour is governed by
the zero-temperature fixed point in the renormalization group equations.
\cite{Mi84,Br87}. Up to now it seems that the Parisi solution to mean-field
theory is essentially correct. It has passed the stability analysis
\cite{Do83} and gives also a correct description of the thermodynamics,
in agreement with the numerical simulations.  It is not clear what is
the correct description of the spin-glass phase in short-ranged models.
Droplet models are expected to be a good description of the low
temperature phase mainly in the case of low dimensions. But droplet
models are not suited to describe the physics of high dimensional
systems and particularly mean-field theory.

The complexity of the replica approach is found when studying the
spectrum of fluctuations around the Parisi solution.  The full set of
gaussian propagators has revealed a very complex structure \cite{Do84}
and the obtention of the one-loop corrections to the mean-field
equations makes progress slow.  The main difficulty of this task is the
enormous number of sectors within replica space which contribute to the
one-loop correction.  This explains also why finite-size corrections to
the main thermodynamic functions are still unknown in mean-field theory
\cite{Sl93}. To all these problems should be added also the fact
that, up to now, the major part of the computations have been done only
close to $T_c$ within the Parisi approximation.

In this work we will try to introduce a different approach to the
problem which can help in understanding the nature of the spin-glass
phase. The main idea of the approach is to try to look for one order
parameter whose spectrum of fluctuations is easier to take into account.
The chaos problem was proposed some time ago and concerns the chaotic
nature of the spin glass phase \cite{Ka82,Br84}. The term chaotic can be
misleading since it can evoke different meanings. In this context, we
prefer to use the word static chaos. By this we mean that a small
perturbation of the Hamiltonian is enough to reshuffle the Boltzmann
weights of the different equilibrium configurations. One constructs a
system which is the sum of two Hamiltonians, the initial plus the
perturbed one. The full system lives in a larger phase space and allows
for a new order parameter. This order parameter is the overlap between
the equilibrium configurations of the initial system and the perturbed
one. This new order parameter has a longitudinal spectrum of
fluctuations without zero modes and hence is stable. The associated
correlation functions to this order parameter decay exponentially to
zero with a caracteristic correlation length.

The nature of the chaos problem is also interesting concerning numerical
techniques like simulated annealing where the the change of the
temperature has to be considered as a perturbation to the system. In
this case one wants to reach the ground state after a progressive
cooling of the system. Let us suppose that the spin glass behaves
chaotically against temperature changes. Then, the equilibrium
configurations should reorganize completely for any small change of the
temperature and a slow cooling would be useless. A small change of
temperature would be considered like a new quenching and the system
would be always strongly far from equilibrium.  Fortunately, as we will
discuss later for this particular problem, if there is chaos against
temperature changes then it is small and finite-size corrections ensure
that a high degree of correlation between the equilibrium configurations
at the two temperatures is preserved.

The work is divided as follows.  In the following section we will
present a quantitative definition for chaoticity and we will introduce
different type of chaotic perturbations.  We will also present
predictions from mean-field theory and phenomenological droplet models.
Section 3 is devoted to the study of a particular perturbation, i.e.
chaoticity against changes of magnetic field. Starting from the
mean-field approach we propose a scaling behaviour within the spin glass
phase. We discuss also our predictions in the framework of the droplet
approach. Section 4 is devoted to numerical simulation results and
finally we present our conclusions and a discussion of the results.

\section{A working definition for chaos}

The idea underlying the chaoticity of the spin-glass phase relies on the
fact that it is a marginal phase \cite{Do84}. The fact that the
spin-glass phase is not fully stable means that a small addition of
energy to the system is able to change completely the statistical
weights of the equilibrium states with a very small cost of free energy
(of order $1/N$ compared to the supplied energy to the system, where $N$
is the size of the system).

Marginality is one of the outstanding results in mean-field theory of
spin glasses. It is also a feature of phenomenological droplet models
and in general it is related to the fact that in the spin-glass phase
spatial (time) correlation functions decay very slowly with distance
(time). This decay is not far from a power law in the most general case.
The full reorganization of equilibrium states in spin glasses after a
small perturbation is a natural feature in mean-field theory. In this
case there is an infinite number equilibrium states and all of them
contribute to the partition function but with a different weight
\cite{Me84}. This is because they have equal free energies per site
except differences of order $1/N$. Any small but finite addition of
energy to the system is enough to redistribute these small free energy
differences reshuffling the weights of the different pure states.

In droplet models there exists the concept of overlap length (it is
sometimes denoted $L_{\Delta T}$ or $L_{\Delta H}$ according to the case
if the perturbation is a change of the temperature or the magnetic
field). Droplet models suppose that there is only one equilibrium state.
When the system is perturbed, the correlation functions reorganize
completely in a scale of distances larger than the characteristic
overlap length. It is clear anyway that the overlap length in these
models has to be always smaller than the correlation length and coincide
only in the limit in which the perturbation vanishes.

In what follows $<...>$ and $\overline{(.)}$ mean thermal and disorder
average respectively. Now we want to give an appropiate definition of
what is static chaos. For simplicity we will consider Ising spin glasses
even though the definition can be generalized to other models.
Let us suppose an Ising spin glass system with Hamiltonian
$H_1[\sigma]$. Then we apply a perturbation $P$ to the system and the
new Hamiltonian for a different copy of spins $\lbrace\tau_i\rbrace$ is
given by
\be
H_2[\tau] = H_1[\tau]+P[\tau]
\ee

We consider now a full Hamiltonian which is the initial system
$H_1[\sigma]$ plus the perturbed one $H_2[\tau]$, i.e.
$H[\sigma,\tau]=H_1[\sigma]+H_2[\tau]$. The phase space has been
enlarged and we can consider a new order parameter which, for example, in
the case of Ising spin glasses, is given by the overlap $\langle
\sigma_i\tau_i\rangle$ between the
equilibrium configurations of the system $H_1$ with the equilibrium
configurations of the perturbed one $H_2$.  When there is no
perturbation, i.e. $P=0$, this is the usual order parameter of the spin
glass with Hamiltonian $H=H_1+H_2$. 

We now define the chaoticity parameter $r$ by:
\be
r(P)=\frac{\overline{\langle\sigma_i\tau_i\rangle_H^2}}{(\overline
{\langle\sigma_i^a\sigma_i^b\rangle_{H_1}^2}\,\,{\overline
{\langle\tau_i^a\tau_i^b\rangle_{H_2}^2})^{\frac{1}{2}}}}
\label {eqr}
\ee
where $\langle\sigma_i^a\sigma_i^b\rangle$ denotes the order parameter
evaluated taking two copies $a,b$ of the unperturbed system $H_1$ and
similarly for $\langle\tau_i^a\tau_i^b\rangle$ of the perturbed system
$H_2$. The thermal average of the order parameter in the numerator is
performed with the full Hamiltonian $H$. In principle, this order
parameter is equal to one if the perturbation is zero. This is trivial
because $r$ is the order parameter of the spin glass normalized to
itself.  Chaoticity in spin glasses reflects the fact that any small but
finite perturbation $P$ causes the parameter $r$ to fall abruptly to a
value smaller than $r=1$. Obviously this can only happen in the
thermodynamic limit because, for a finite size $N$, the chaos parameter
$r$ will always be a smooth function of the perturbation $P$. This means
that one has to perform the thermodynamic limit before applying the
perturbation $P$.  More precisely, the spin-glass phase is chaotic if
\be
\lim_{P\to 0}\,\lim_{N\to\infty}\,r(P) < 1
\label{eqrN}
\ee

It is also possible to define the adimensional quantity 
\be
a=\frac{\overline{\langle \sigma_i\tau_i\rangle_H^2}}
{\overline{\langle \sigma_i^a\sigma_i^b\rangle^2}_{H_1}}
\label{eqa}
\ee
where the numerator is obtained by averaging over the Hamiltonian
$H[\sigma\tau]$ and the denominator is the order parameter for two
copies $a$ and $b$ of the same system $H_1[\sigma]$. The difference
between the adimensional quantities $a$ and $r$ is only an appropiate
normalization. In fact, the definition of chaos given above in
eq.(\ref{eqrN}) also holds in case of the parameter $a$. The necessity
to distinguish among the parameter $a$ and the parameter $r$ is
important for certain types of perturbations.  For example, in the case
of temperature changes the order parameter
$\langle\sigma_i^a\sigma_i^b\rangle$ is very sensitive to the
temperature and vanishes at the critical point. Let us suppose the
initial Hamiltonian is in the low temperature phase and we change the
temperature by putting the system close to the critical point. The chaos
parameter $a$ vanishes because the numerator in eq.(\ref{eqa}) vanishes
close to $T_c$ and the denominator remains finite. On the contrary, the
chaos parameter $r$ of eq.(\ref{eqr}) has in the denominator a term
which also vanishes at $T_c$ and normalizes appropiately the numerator.
Because $r$ measures correctly the overlap among equilibrium
configurations it is the appropiate parameter to deal with in case of
temperature changes. The difference between the chaos parameters $a$ and
$r$ is not important in the case of magnetic field changes and other types
of perturbations.

Let us now discuss what happens in the case of ordered systems. As an
example we take the standard Ising model in a finite number of
dimensions. Let us suppose that we are in the low temperature phase, at
temperature $T$ below the critical point, and let us take as a
perturbation a small change of the temperature. At the temperature $T$
the system has a spontaneous magnetization $m$. There is only one
equilibrium configuration with a fraction $m$ of the spins pointing in a
certain direction. When we change a little bit the temperature by a
small quantity $\Delta T$, the mean number of spins which point in that
direction (i.e. the magnetization) changes linearly with $\Delta T$ at
least for $\Delta T$ small. In this case one sees inmediately that
eq.(\ref{eqr}) gives the value $r(\Delta T)=1$ for any small change of
the temperature and the system is not chaotic. We represent an
equilibrium state by an $N$ dimensional vector $v=(m_1,m_2,...,m_N)$
where the $m_i$ are the local spin magnetizations.  In the Ising model a
slight change of the temperature modifies only the length of this vector
but not its direction. The chaos mechanism in spin glasses is driven by
the fact that as soon as we perturb the system this vector $v$ suffers a
sudden rotation because the weight of the different states are changed.
In most cases, any small perturbation makes the vector $v$ to become
orthogonal to its previous value and $r(P)$ vanishes for a finite
perturbation $P$.

There are many examples of perturbations that one can apply to the
system. As was mentioned in the introduction, one can change the
temperature or change the magnetic field. These are among the most
studied perturbations in the literature. But one can imagine other kinds
of perturbation like for instance changing the realization of disorder.
In this case, a finite fraction of the $J_{ij}$ couplings is changed
(for instance, in case of symmetric distribution of couplings, this
change could consist in reversing the sign of the perturbed couplings
$J_{ij}$). One can also imagine to add a small ferromagnetic or
antiferromagnetic part to the couplings.  In these cases, spin glasses
seem to behave chaotically against these perturbations.  As an example,
we show in figure 1 how the parameter $r$ decreases with the size $N$,
for the case of the SK model for two different perturbations. The first
perturbation corresponds to a very large change of temperature ($\Delta
T=0.4$) for an initial temperature $T=0.5=T_c/2$ (the perturbed system
is at $T=0.9$ which is very close to $T_c$ but always in the spin-glass
phase.) The other perturbation is the application of a small magnetic
field $h=0.2$ to a system initially at zero magnetic field and $T=0.6$
(the AT line lies at $h\sim 0.4$). The results for $N$ less than 20 have
been obtained by calculating exactly the partition function, the
remaining ones using Monte Carlo simulations.  From figure 1, the system
seems much more sensitive against magnetic field perturbations than
temperature changes.  This is clear also if we observe that under the
temperature perturbation, even though it is very strong because it puts
the perturbed system close to the paramagnetic phase, the equilibrium
configurations at both temperatures still retain a high degree of
coherence $(r\sim 0.7)$.

In the following, we will focus on the study of a particular
perturbation, which has turned out manageable in order to understand its
effects in the spin glass phase: the case in which the perturbation
consists in applying a small magnetic field to a spin-glass at zero
field.  This has been the subject of previous research, specially by I.
Kondor in the case of mean-field theory \cite{Ko89}. One could also
study the case in which the system is at a finite field in the
spin-glass phase and the field is slightly changed. This problem is more
subtle than the previous one in which the system is initially at zero
magnetic field. The main reason is that (at least for short ranged
systems) we do not know if the spin-glass phase survives to a magnetic
field. If the spin-glass phase survives to the magnetic field then we
expect (as predicted in the mean-field approach) that chaoticity will be
present in a magnetic field. In the other case (and this is the
prediction of droplet models), the system would then be always in the
paramagnetic phase and chaos should not be present. Then, according to
eq.(\ref{eqr}), $r(P)=1+O(\Delta h)$ would be continuous for $\Delta
h=0$.

In the case the system is initially at zero magnetic field mean-field
approach and droplet models agree in that they both predict that the
spin-glass phase is chaotic. More specifically $r(h)$ (we use the
intensity of the applied magnetic field $h$ for the pertubation $P$) is
zero for any finite $h$.  But the main mechanism which makes the
spin-glass phase chaotic is very different in both pictures. In
phenomenological droplet models the spin-glass phase is marginal: the
correlation functions decay very slowly with the distance and the
correlation length associated to the two point function
$C(x)=\overline{\langle\sigma (0)\sigma (x)\rangle^2}$ is infinite. When
a magnetic field is applied the spin-glass phase is destroyed and the
correlation length becomes finite. It is given by \cite{Fi88}:
\be
\xi\sim (q_{EA}\,h^2)^{\frac{1}{2\theta-d}}
\label{xih}
\ee
with $q_{EA}$ the Edwards-Anderson order parameter and $\theta$ the
thermal exponent which gives the characteristic energy scale
$L^{\theta}$ of droplet excitations of typical size $L$. This means that
all excitations of droplets of sizes larger than a certain length $\xi$
will be suppressed by the field.  The exponent $\theta$ is a zero
temperature exponent (it is determined by the zero temperature fixed
point of the renormalization group equations) and it is expected to be
constant in the low temperature phase.  In the critical point the
associated thermal exponent $\theta_c$ is determined by the finite
temperature fixed point of the renormalization group equations and is
related to the critical exponents by $\theta_c=\frac{d-2+\eta}{4}$ where
$\eta$ is the anomalous dimension exponent an $d$ is the dimension (even
though it has been argued that at low dimensions there appears a new
exponent $\theta_c$ \cite{Ni92}).  In general, we expect that $\theta_c$
is smaller than $\theta$ above the lower critical dimension and both
vanish in the lower critical dimension.

Mean-field theory approach gives a completely different mechanism of
chaoticity.  After applying a small magnetic field, the spin-glass phase
is not destroyed.  We suppose that the effect of the magnetic field is
the dissapareance of a large number (infinite) of equilibrium states.
This mechanism is easy to visualize by taking into account the correct
order parameter for spin glasses which is the distribution $P(q)$. Its
physical meaning was explained some time ago \cite{Doyo83,Pa83} and it
gives the probability density that two pure states $\alpha$ and $\beta$
have a common overlap $q_{\alpha\beta}=q$.  This common overlap
corresponds to the scalar product of the local spin magnetization in
both states. At zero magnetic field the function $P(q)$ is symmetrically
distributed around $q=0$ and non zero within the interval
$(-q_{max},q_{max})$.  In a magnetic field the reversal symmetry
$\sigma\to-\sigma$ is broken and $P(q)$ is non zero only for $q$
positive and larger than a minimum value $q_{min}$. The value of
$q_{max}$ is nearly independent of the magnetic field (this
approximation, which works extremelly well close to $T_c$, is called the
Parisi-Toulouse hypothesis \cite{To80}). In some sense the effect of the
magnetic field is to suppress those equilibrium states $\alpha$ which
had overlaps $q_{\alpha\beta}$ with the other remaining states $\beta$
smaller than $q_{min}$. Within the usual picture of the spin-glass phase
in mean-field theory \cite{Me84} there is an infinity of states with a
few number of them dominating the Gibbs measure. This infinity of states
lay in the tips of an ultrametric tree and the effect of the magnetic
field corresponds to progressively cutting those branches which generate
the states which are suppressed. The suppression of the states also
conserves the ultrametricity property.  The understanding of how pure
states $\alpha$ are suppressed by the field according to their
statistical weight $w_{\alpha}$ is still an interesting open problem.

\section{Chaos in magnetic field}

This section is devoted to the study of chaos in case of an Ising spin
glass initially at zero field after turning on a magnetic field.  We are
interested in the case of a $d$-dimensional Ising spin glass with random
$J_{ij}$ couplings defined by the Hamiltonian
\be
H=-\sum_{(i,j)}\,J_{ij}\sigma_i\sigma_j\,-\,h\sum_i\sigma_i
\label{eqH}
\ee
where the couplings $J_{ij}$ are quenched variables distributed
according to a probability function $P(J)$ of zero mean and finite
variance. The interaction is restricted to nearest neighbours and $h$ is
the magnetic field. The Ising spins $\sigma_i$ take the two possible
values $\pm 1$ and live in a $d$-dimensional hypercubic lattice. In the limit
$d\to\infty$ one expects to converge to mean-field theory, i.e. the SK
model.  In the SK model all spins interact one to each other
and the couplings $J_{ij}$ are normalized by a factor $1/\sqrt{N}$ where
$N$ is the number of spins. 

\subsection{The case of mean-field theory}

This question was adressed some time ago by I. Kondor \cite{Ko89}. Let
us consider two copies of the same realization of disorder, one at zero
magnetic field and the other one at finite magnetic field $h$. The full
Hamiltonian of the problem is given by:
\be
H[\sigma\,,\tau]=H_1[\sigma]+H_1[\tau]- h\sum_i\tau_i
\label{eq2}
\ee
with
\be
H_1=-\sum_{1\leq i<j\leq N}\,J_{ij}\,\sigma_i\,\sigma_j 
\label{eq3}
\ee
This problem can be directly solved using the standard replica trick for
the full Hamiltonian $\sum_{a=1,n}H_a$ where $H_a$ is given by
eq.(\ref{eq2}) and $a$ is the replica index which runs from 1 to the
full number of replicas $n$ (at last one takes the limit $n\to 0$). Now
one applies the replica trick
\be
\overline{\log Z}=\lim_{n\to 0}\frac{\overline{Z_J^n}-1}{n},
\label{eq4}
\ee
which yields the expression
\be
\overline{Z_J^n}=\int dP_{ab}dQ_{ab}dR_{ab} \exp (-NA[PQR])
\label{eq5}
\ee
with
\begin{eqnarray}
A[PQR]&=&\sum_{a<b}(P_{ab}^2+Q_{ab}^2+2R_{ab}^2)+\sum_a\,R_{aa}^2- \log
Tr_{\sigma\tau}\,\exp\Bigl
(\beta^2\sum_{a<b}P_{ab}\sigma_a\sigma_b\,\nonumber\\
&&+\,\beta^2\sum_{a<b}Q_{ab}\tau_a\tau_b\,+\,\beta^2\sum_{a\neq
b}R_{ab}\sigma_a\tau_b\,+\,\beta^2\sum_aR_{aa}\sigma_a\tau_a\,\nonumber\\
&&+\,\beta h\sum_a\tau_a)
\end{eqnarray}

In this way, one is able to reduce the problem in terms of a lagrangian
$A[P\,Q\,R]$ with three order parameters corresponding to the different
overlaps among the two copies, i.e
$P_{ab}=\langle\sigma_a\sigma_b\rangle$; $P_{aa}=0$ ,
$Q_{ab}=\langle\tau_a\tau_b\rangle$; $Q_{aa}=0$ and
$R_{ab}=\langle\sigma_a\tau_b\rangle$. For finite $h$ there is an
inmediate solution for the equations of motion which is given by the
$P_{ab}$ and $Q_{ab}$ Parisi matrices with zero and magnetic field $h$
respectively and $R_{ab}=0$. The free energy of the whole system is the
sum of the free energy of one copy at zero magnetic field plus the free
energy of the other copy with field $h$. This is a solution
because it gives the full free energy of two uncoupled systems. The
order parameter associated to $R$ is
\be
q=\sum_i\sigma_i\tau_i
\label{op}
\ee
In order to study the stability of this solution one computes the
spectrum of fluctuations. The full set of fluctuations is very
complex. For instance, within the subspaces generated by the diagonal
subblocks $P$ and $Q$, it corresponds to the
spin-glass spectrum derived by C. De Dominicis and I. Kondor
\cite{Do84}. Only the fluctuations around $R=0$ (the off-diagonal
subblock) are those which are physically relevant to the problem because
they measure spatial correlations between states corresponding to the
two Hamiltonians, the initial and the perturbed one.  In mean-field
theory there are no distances and we want to obtain the spatial
behaviour of the system within the mean-field approximation.  This can
be done using a Ginzburg-Landau approximation by introducing spatially
dependent order parameters in the effective action of eq.(\ref{eq5}).
Now the order parameters $P,Q,R$ depend on the space variable $x$ and we
add to the action $A$ a kinetic term of the type
$\sum_{a<b}\frac{\partial^2 R_{ab}(x)}{\partial x^2}$.  The spectrum of
fluctuations is contained in the the momentum space propagator. This is
given by the Fourier transform $G(p)$ of the correlation function
\be
C(x)=\overline{\langle\sigma_0\sigma_x\tau_0\tau_x\rangle}
\label{eq7}
\ee
where $\overline{(\cdot)}$ means averaging over disorder and
$\langle(\cdot)\rangle$ is the usual thermal average over the
Hamiltonian eq.(\ref{eq2}).

The problem of computing the propagator reduces to the diagonalization
of a hierarchical matrix of the Parisi type. The full expression has been
reported in \cite{Ko89}. Its singular part is given by
\be
G(p)=\int_{0}^{q_{max}}\,dq\,\int_{Q_{min}}^{Q_{max}}\,dQ
\frac{p^2+1+\lambda(q)\lambda(Q)}{(p^2+1-\lambda(q)\lambda(Q))^3}
\label{eq8}
\ee
with
\be
\lambda(q)=\beta(1-q_{max}+\int_{q}^{q_{max}}\,dq\,x(q))
\label{eq9}
\ee
where $\beta$ is the inverse of the temperature. The same expression
applies in the case of $\lambda(Q)$. Here $q(x)$ and $Q(x)$
are the order parameter functions for the spin glass at zero and $h$
field respectively.

The correlation function eq.(\ref{eq7}) decays to zero for large
distances $x$ with a characteristic length $\xi$ which is given by the
minimum eigenvalue of the stability matrix. This eigenvalue is non-zero
for finite $h$ which demonstrates the stability of the $R=0$ solution.
The correlation length $\xi$ (which is the inverse square root of the
minimum eigenvalue) diverges like $(1-\lambda(Q_{min}))^{-\frac{1}{2}}$.
Close to $T_c=1$ we have $Q_{min}\sim h^{\frac{2}{3}}$. This gives
$\xi\sim h^{-\frac{2}{3}}$ which diverges when $h\to 0$.

The stability of the $R=0$ solution implies that the system is chaotic.
This means that $r(P)$ (as given by eq.(\ref{eqr})) always vanishes for
finite $h$ like $\frac{1}{N}$, where $N$ is the size of the system. The
result that $\xi$ diverges when $h$ goes to zero is rather natural
because in that case the perturbation vanishes and the two copies are
identical. Then the correlation length $\xi$ is the spin-glass correlation
length which is infinite because there is marginal stability. In some
sense, there is a first order phase transition at $h=0$ where the
probability distribution associated to the order parameter $q$ defined
in eq.(\ref{op}) changes from a delta function peaked in $q=0$ to the
usual order parameter distribution for the spin glass \cite{Yo83}. Also
the critical point is chaotic but in this case the correlation lenth
diverges like $\xi\sim h^{-\frac{1}{2}}$.  We remind the reader that in
the paramagnetic phase there could not be chaos of the type defined in
eq.(\ref{eqrN}). This is because the correlation length $\xi$ would
never diverge but only converge smoothly to its corresponding finite
value at zero magnetic field.

Now we turn to the behaviour of the propagator $G(p)$ of eq.(\ref{eq8})
in the limit $p\to 0$. Using the known expressions
\cite{Pa80} for $q(x)$ and $Q(x)$ close to $T_c$ in eq.(\ref{eq8}) we
obtain a divergent expression for $G(0)$. Its most divergent part is
\be
G(0)\sim
\int_{Q_{min}}^{Q_{max}}\,\frac{dQ}{(1-\lambda(Q))^{\frac{5}{2}}}
\label{eq10}
\ee
which gives $G(0)\sim p^{-4}\sim\xi^4\sim h^{-\frac{8}{3}}$.
This is not new and this result can also be obtained from the study of
the intravalley gaussian propagators as derived in \cite{Te88}.
We can define a certain kind of non-linear susceptibility by:
\be
\chi_{nl}=\sum_i C(i)=G(0)\sim h^{-\frac{8}{3}}
\label{eq11}
\ee
This susceptibility can also be written $\chi_{nl}=N\,\overline{\langle
q^2\rangle} $ with $q$ given in eq.(\ref{op}).  Using the fact that
$R=0$ is a stable solution altogether with eq.(\ref{eq11}), the
following scaling behaviour holds
\be
a\equiv r\sim f(N h^{\frac{8}{3}})
\label{eq12}
\ee
This result will be derived in the following section using scaling
arguments and will be also generalized to short-range models.

\subsection{Scaling theory of chaos with magnetic field}

Next we want to give a precise physical meaning to the correlation
length $\xi$. As commented in the previous section, the spin glass phase
is marginal with an infinity of equilibrium states, none of them having
a characteristic correlation length. Under a small magnetic field, a lot
of states are suppressed and the correlation length $\xi$ is finite.  We
interpret $\xi$ as the new typical correlation length of the states
which have been suppressed. This is the natural continuation of what
comes out in the critical point. In this case there is only one marginal
state. After applying a magnetic field, the correlation length becomes
finite and the system goes into the paramagnetic phase.  In the
spin-glass phase the suppressed states acquire finite correlation length
and their free energy increases respectively to the remaining ones. We
are still within the spin-glass phase because the remaining states
dominate the partition function and they still have infinite correlation
length.  In the spin-glass phase all equilibrium states are non
equivalent. Some of them have a much higher statistical weight. This
means that only those states $\lbrace\alpha\rbrace$ which give overlaps
$\lbrace q_{\alpha\beta}\le q_{min}\,\forall\beta\rbrace$ are simply
erased by the magnetic field. When all states are suppressed we reach
the AT line \cite{At78}.  This can only happen in case when an
infinity of equilibrium states coexist in the low temperature phase.

If we want to be more precise we have to generalize these ideas to the
case of short-range models. Two basic assumptions are enough to this
aim. The first one concerns the physical interpretation on the effect of
the magnetic field on the equilibrium states. The second one uses
information in the critical point to understand what happens in the
spin-glass phase. More precisely, we suppose that the cost in free
energy of the dissapearing states in the spin-glass phase, scales in the
same way as in the critical point. We argue that the low temperature
spin-glass phase is determined by the finite temperature fixed point of
the renormalization group equations. This is the contrary assertion of
droplet models in which the spin-glass beahviour is governed by the
zero-temperature fixed point. Our assumptions give exact
results in mean-field theory. The existence of some critical properties
in the low temperature phase has been also seen in a different context.
For example, it has been proved that the exponent which characterizes
the decay of the tail of the $P(q)$ around $q=q_{max}$ freezes below the
critical point in mean-field theory \cite{Sl93}. In short-range Ising
spin glasses there are also numerical results which suggest that the
freezing of some critical exponents really takes place in the low
temperature phase
\cite{Pa93,Ci93}.

At the critical point we know that the singular part of the free
energy (per site) is given by
\be
f_{sing}\sim h^2 q\sim q^{\frac{d}{[q]}}
\label{eq13}
\ee
Here $q$ is the usual order parameter defined in eq.(\ref{op}), $d$ is
the dimension and $[q]$ is the dimension of the operator $Q_{ab}$ in
units of the inverse of the correlation length. the value $[q]$ is
connected to the critical exponents $\beta,\nu$ and $\eta$ via the
relation $[q]=\frac{\beta}{\nu}=\frac{d-2+\eta}{2}$.

Now we generalize this expression to the case in which replica symmetry
is broken, i.e. in the spin-glass phase. First of all, we need a general
expression for the singular part of the free energy which is invariant
under the permutation group of the different replicas. The most easy
expression of this type is
\be
f_{sing}\sim\sum_{a<b}\,Q_{ab}^{\frac{d}{[q]}}
\label{eq14}
\ee
where the exponent $[q]$ is given by the critical exponents.

One can easily derive the correct behaviour of the singular part of the
free energy for an order parameter $q(x)$ of the type shown in figure 2.
To obtain the correct singular part of the free energy corresponding to
the states which are suppressed by the field we have to take the
difference of eq.(\ref{eq14}) with $h\neq0$ and $h=0$
\begin{eqnarray}
f_{sing}&=&\sum_{a<b}\,Q_{ab}^{\frac{d}{[q]}}(h)-\sum_{a<b}\,
Q_{ab}^{\frac{d}{[q]}}(0)=\nonumber\\
&&\int_0^1\,(q^{\frac{d}{[q]}}_h(x)-q^{\frac{d}{[q]}}_0(x))\,dx=
x_{min}\,q_{min}^{\frac{d}{[q]}}= q_{min}^{\frac{d}{[q]}+1}
\label{eq15}
\end{eqnarray}

The main ingredient that we have used in this derivation is the fact
that the order parameter $q(x)$ in short-range models is characterized
by a continuous part plus a plateau. Under the application of a magnetic
field a new plateau appears with $q(x)=q_{min}$ and $x_{min}\sim
q_{min}$. This last result is connected with the fact that the order
parameter distribution $P(q)$ at zero magnetic field is finite for
$q=0$. Because $P(q)=\frac{dx(q)}{dq}$ (where $x(q)$ is the invers of
the $q(x)$) this means that $q(x)\sim x$ for $x$ close to zero. In the
critical point the previous derivation applies with $x_{min}=1$ and
$q\sim h^\frac{2}{\delta}$ where $\delta=\frac{d+2-\eta}{d-2+\eta}$. For
droplet models the same derivation is valid but now $x_{min}=1$ and
$P(0)$ vanishes in the thermodynamic limit like $L^{-\theta}$ with
$\theta$ the zero-temperature fixed point exponent. The singular part of
the free energy scales like $\xi^{-d}$, $\xi$ being given by
eq.(\ref{xih}).

Now we apply eq.(\ref{eq15}) to mean-field theory. Mean-field critical
exponents together with $\eta=0$ give $[q]=2$. Because $q_{min}\sim
h^{\frac{2}{3}}$ we obtain $f_{sing}\sim h^{\frac{8}{3}}$ and the global
singular free energy scales like $N\,h^{\frac{8}{3}}$.  Because the
parameters $a$ and $r$ are adimensional we reproduce the scaling
behaviour of eq.(\ref{eq12}).

\subsection{Estimate of the AT line and the lower critical dimension}

The first result which comes out from the previous subsection is that
$d=4$ plays a role as a special critical dimension. This deserves some
explanation. The upper critical dimension in Ising spin glasses is $6$.
Above 6 dimension the critical exponents coincide with the mean-field
ones. These exponents are associated with the order parameter $Q(x)$
corresponding to the overlap $\sigma^a(x)\sigma^b(x)$ between two copies
$a,b$ with the same Hamiltonian. The correlation length associated to
the two point function $\langle Q(0)Q(x)\rangle$ diverges at the critical
point and remains infinite in the low temperature phase. The chaos
correlation length is associated to the two point function
$\langle R(0)R(x)\rangle$ and corresponds to a different order parameter
$R(x)=\sigma(x)\tau(x)$ which couples two systems with different
Hamiltonians.  We argue that the exponent of the chaos correlation length
$\xi$ associated to $R(x)$ lies in a different universality class of
that to which the order parameter $Q(x)$ belongs.

We can find the appropiate upper critical dimension associated to the
criticality of chaos. From eq.(\ref{eq12}) and using $\xi\sim
h^{-\frac{2}{3}}$ we obtain an argument of the scaling function for $a$
of the form $L/\xi$ in four dimensions. This means that $d_u=4$ has the
role of an upper critical dimension.

In the most general case we can introduce the exponent $\lambda$ defined
by $q_{min}\sim h^{\frac{2}{\lambda}}$. We expect the following scaling
to be satisfied
\be
a\equiv r\sim f(N\,h^{\frac{2(d+[q])}{\lambda [q]}})
\label{eq16}
\ee
The value $[q]$ depends on the critical exponents and this scaling
contains only one non critical parameter ($\lambda$) and thus is easily
measurable in a simulation.

The exponent $\lambda$ is theoretically unknown and there is no
numerical prediction on its value. From the value of $\lambda$ one
obtains the AT line using the condition $q_{min}\sim q_{max}$. Because
$q_{max}\sim\tau^{\beta}$ with $\tau=\frac{T_c-T}{T_c}$ the AT line is
given by the equation $h\sim\tau^{\frac{\beta\lambda}{2}}$. In
mean-field theory $\beta=1$ , $\lambda=3$ gives
$h\sim\tau^{\frac{3}{2}}$.  In $d=4$ depending on the value of $\lambda$
and $\beta$, a different expression is found. This and the special case
$d=3$ are left as a discussion in the next section.

One can also estimate what is the lower critical dimension $d_l$. In
fact, we expect that the scaling eq.(\ref{eq16}) for $T_c=0$ should be
of the form $a\equiv f(N\,h^2)$. This gives $d+[q]=\lambda [q]$, i.e.
$[q](\lambda-1)=d$. The exponent $\lambda$ should diverge as $d$
approaches $d_l$ because $q$ is discontinuous at $T_c=0$ when a magnetic
field is applied. So $[q]=0$, i.e, $d_l-2+\eta=0$ which is the usual
relation determining the lower critical dimension \cite{Do93} (in
principle this relation should at least be satisfied for Hamiltonians
with a countinuous distribution of couplings). Furthermore, in case
$d=d_l$ one expects $\xi\sim h^{-\frac{2}{d_l}}$. Because $\xi\sim
h^{-\frac{2}{3}}$ for $d=d_u=4$ this means that $d_l=3$ at zero order of
approximation or mean-field level. We call it mean-field level or zero
order because in this case we suppose the exponent for the correlation
length $\xi$ does not vary between $d_u=4$ and $d=3$. The fact that 4
and 3 are very close assures that this is a good approximation which is
probably exact.

We should now recall that all these predictions have to be apropiately
modified for droplet models. For these models there is no transition in
a magnetic field. The condition $d_l-2+\eta=0$ also applies and the
exponent $\theta$ is well approximated by the result
$\theta=\frac{d-3}{2}$.  This is in agreement with the numerical results
of the following section. Recent numerical simulations for case $d=2$
show that the zero temperature exponent $\theta\sim -0.46$ \cite{Ta92}
is surprisingly close to the chaos prediction $-0.5$. This suggests that
in the framework of droplet models also $d_l=3$ and the previous
expression for $\theta$ are probably exact.

\section{Numerical results}

In this section we present Monte Carlo numerical simulations in order to
test these ideas. We should note that the chaos parameters $a$ and $r$
defined in this work and all the scaling laws based on them are
computable using standard numerical simulations. The standard technique
is to consider two parallel Monte Carlo simulations, one for the system
$H_1$ and the other one for the perturbed system $H_2=P[H_1]$. The first
copy is at zero magnetic field while the second one has a magnetic field
$h$.  Both copies evolve in time and, once they have thermalized, one
computes the corresponding order parameters. Since one is interested in
scalings within the spin-glass phase, the main difficulty is that
samples have to be equilibrated in the low temperature phase where
metastability is very strong. All the results in this section are for
small lattices and we have paid attention that they are fully
equilibrated. The general schedule of the simulation is as follows. An
initial cooling is performed until the first copy at zero magnetic field
thermalizes at the working temperature and the second copy thermalizes
with an applied magnetic field equal to a maximum value $h_{max}$. Then,
the first copy evolves without any perturbation and the field of the
second copy is progressively decreased step by step down to zero.  In
general, for each different value of the magnetic field of the second
copy, a long enough thermalization is done after which statistics is
collected. Then, the order parameters $a$ and $r$ of eqs.(\ref{eqa}) and
eq.(\ref{eqr}) can be computed.

The Hamiltonian under study is given by eq.(\ref{eqH}).  In all
simulations we have used the heat-bath algorithm and spins are updated
sequentally. In case of short-range models we impose periodic boundary
conditions. The distribution of the coupling $J$ is discrete (the
$J_{ij}$ can take the values $\pm 1$ with equal probability).  If there
is a finite temperature phase transition we expect universality to apply
(anyway see \cite{Ca94}) and the results for discrete couplings should
be equivalent to the case in which the distribution is continuous.  Our
main goal is now to test scaling laws of the type eq.(\ref{eq16}).
Scaling fits work well if we use the parameter $r$ or the parameter $a$
(in all cases they differ very slightly, approximately by 5 per cent).
Then, we will present results only for the parameter $a$.

We now show the results in case of mean-field theory.  The results for
different magnetic fields ranging from $0.2$ up to $1.0$ at $T=0.6$ are
shown in figure 3 for several sizes. We show the parameter $a$ versus
$h\,N^{\frac{3}{8}}$ (we have chosen this argument instead of
$N\,h^{\frac{8}{3}}$ in order to compare directly these mean-field
results with those corresponding to short-range models.) There is an
agreement with the prediction of eq.(\ref{eq12}). At the critical point
$T=1$ the appropiate scaling argument is $N\,h^3$ and in order to
compare with the scaling law eq.(\ref{eq12}) of figure 3 we show results
for $T_c=1$ in figure 4.  If in figure 3 we plot the chaos parameter $a$
in function of $N\,h^{\frac{8}{3}}$ (instead of $h\,N^{\frac{3}{8}}$) we
discover that the scaling functions of figures 3 and 4 are clearly
different suggesting that the criticality of chaos in the critical point
and in the low temperature phase are in a different universality class.

Next we present results for the case $d=4$. Figure 5 shows the parameter
$a$ as a function of $L\,h^{\frac{2}{3}}$. This is the mean-field
scaling which is in full agreement with data.  Simulations were
performed at $T=1.5$ ($T_c\simeq 2.05$ \cite{Si86}) which is $\simeq 0.7
T_c$.  Metastability effects are very strong and thermalization is more
difficult (in the sense that one needs more thermalization steps) than
in the SK model case. Error bars are not shown because they are very
small (of order of the size of the symbols). The agreement with the
theoretical prediction is good. 

Then we can derive results for the AT line. Using eq.(\ref{eq16}) we get
$\lambda=4.2$ which gives $q_{min}\sim h^{0.48}$. In the critical point
taking $\eta\sim -0.25$ and using the hyperscaling relation
$\delta=\frac{d+2-\eta}{d-2+\eta}$ where $q\sim h^{\frac{2}{\delta}}$
one gets the result $\delta\sim 3.6$. In the critical point, $q$ scales
with the magnetic field with a larger exponent $q\sim h^{0.55}$ and one
has the impression that this is a general feature at any dimension (in
the mean-field case $q$ scales linearly with the field in the critical
point while the minimum overlap scales like $h^{\frac{2}{3}}$ in the
spin-glass phase.) Because the critical exponent $\beta\sim0.6$ and
$\eta=-0.25\pm0.1$ the corresponding AT line should scale like
$h\sim\tau^{1.3\pm 0.1}$ which is close to the mean-field theory result
(even though there is no reason that it should coincide). Unfortunately
we have no means to test if this prediction is correct, mainly because
the question of the existence of the AT line is still unsolved
\cite{Br80,Gr92,Ru93}. In the framework of droplet theory we can derive
the correct value of the zero-temperature exponent $\theta$. It gives
$\theta=0.5$ (in the critical point the finite-temperature exponent
$\theta$ is $\theta_c\sim0.43$). Figure 5 also gives valuable
information in the case there is no AT line. The chaos correlation
length should correspond to the correlation length of the spin-glass in
the paramagnetic phase as given in eq.(\ref{xih}). In the case of $d=4$
we obtain $\xi\simeq 5h^{-\frac{2}{3}}$ if we estimate $\xi$ as the
distance over which the chaos paarmeter $a$ decreases by an order of
magnitude. In order to search numerically for the existence or not of
phase transition in magnetic field at $T=1.5$ one should study lattice
sizes $L>\xi$ where $\xi$ is given by the previous expression.

We analyze now the data for $d=3$. Simulations were performed for the
$\pm J$ nearest-neighbour Ising spin glass. Recent numerical simulations
suggest that there is only a singularity at $T=0$ \cite{Ma93}. But, due
to the so large correlation length, we expect that the system will have
some kind of pseudocritical behavior for small lattices. In fact,
standard numerical simulations for small sizes show that $T_c\sim1.2$
with $\eta\sim-0.1\,,\beta\sim 0.5$ \cite{Bh88,Og85}. This means that,
even if there is no true phase transition in the thermodynamic limit,
simulations for small lattices should be sensitive to the pseudocritical
behaviour and finite-size scaling for the chaos parameter $a$ should
give information regarding this pseudocritical point. What comes out is
very interesting and has been plotted in figure 6. The mean-field result
$\xi\sim h^{-\frac{2}{3}}$ fits very well the data. This is in agreement
with the fact that $d_l=3$ which is the value for the lower critical
dimension if mean-field theory is exact. If $d_l=3$ then any finite $T$
belongs to the paramagnetic phase which is characterized by the true
finite correlation length $\xi_T$ (in order to distinguish it from the
chaos correlation length $\xi$).  The chaos correlation length $\xi$
would behave like $h^{-\frac{2}{3}}$ in the regime $\xi<<\xi_T$.  In the
regime $\xi_T\sim\xi$ the value of the chaos correlation length $\xi$
should progressively match the value of $\xi_T$ and remaining finite.
We can derive in this regime of sizes the location of the pseudocritical
AT line. From eq.(\ref{eq16}) we derive $q_{min}\sim h^{0.26}$ which
gives $h\sim\tau^{1.9}$ for the critical exponent $\beta\sim 0.5$.  This
result is not far from experimental determinations of the AT line in
bulk CuMn spin glasses
\cite{Or93} (where a scaling of the form $h\sim\tau^{1.8}$ with an
exponent slightly larger than the mean-field result $3/2$ is compatible
with experiments.)

\section{Conclusion}

It seems that the study of static chaos in spin glasses can give
interesting predictions of the nature of the spin glass phase. The
information obtainable from the subject is great because of the
different ways one can perturb the system. In this work we have focused
in a magnetic field perturbation. In this case it is possible to
establish a physical picture in which states are supressed by the action
of the magnetic field. By 'suppression of the states' we mean that these
states increase their free energy and do not contribute any more to the
partition function.  Similarly to what happens at the critical point
where there appears a finite correlation length after applying a
magnetic field, these suppressed states acquire a finite correlation
length (the chaos correlation length). To prove this result we should
know what is the real mechanism of the modification of the free energies
of the different equilibrium states.  This means to understand how
states are suppressed by the magnetic field depending on their
statistical weights.  This is an interesting analytical problem in
mean-field theory which is possibly not out of reach.

Using this ideas we have been able to derive a scaling behaviour for the
chaos parameters $a$ and $r$ which depends on the critical exponents at
$T_c$. We argue that some critical exponents survive in the spin-glass
phase which means that the low temperature phase is governed by the
finite-temperature fixed point. Curiously this is the opposite assertion
of droplet models in which the spin-glass behaviour is governed by the
zero-temperature fixed point. The complete understanding of the correct
description of the low temperature phase in spin-glasses is one of the
major still open problems. Mean-field approach yields the value $2/3$
for the chaos correlation length exponent. We expect this exponent to be
exact down to $d=4$ which is the upper critical dimension if
hyperscaling applies for the singular part of the free energy of the
supressed states.  Our numerical results are in very good agreement with
this prediction.  Surprisingly this 'mean-field' behaviour seems exact
down to $d=3$ which should correspond to the lower critical dimension.
In the framework of droplet picture our numerical results suggest that
the relation $\theta=\frac{d-3}{2}$ is probably exact. Both approaches
predict that 3 is the lower critical dimension even though we expect
some kind of pseudocritical behaviour in the regime in which lattice
sizes are smaller than the true correlation length. This pseudocritical
behaviour is expected also with finite magnetic field giving a
pseudocritical AT line. Our results predict a transition line in
agreement with some experimental results.  For $d>d_l=3$ the physics is
determined by the existence or not of a spin-glass phase with magnetic
field. A definite answer on the existence or not of the AT line in
finite dimensions is a prioritary task in order to clarify the
controversy on the real nature of the spin-glass phase (see \cite{Pi94}
for some recent numerical results.)
 
Now we say few words in case of temperature changes. In this case, as
shown in figure 1, chaos is much weaker. This is a interesting result
which finds also a natural explanation in the context of mean-field
theory. This result has already been shown in
\cite{Ko89} doing the same kind of analysis as has been performed in
section 3.1. Namely, when the initial system at temperature $T$ and the
perturbed one at temperature $T'$ both lie within the spin-glass phase then
$\lambda(q)$ in eq.(\ref{eq9}) is always equal to one which corresponds
to an infinite correlation length. This means that if chaos exists then
it is marginal and the chaos parameter $r$ for any finite perturbation
$\Delta T$ goes to zero like $N^{-\alpha}$ with an exponent $\alpha <
1$. Obviously this result does not exclude the possibility that chaos is
absent and $\alpha=0$. Why chaos is much weaker in case of temperature
changes than for magnetic field perturbations can be intuitively
understood if one imagines that by lowering the temperature new
equilibrium configurations emerge from previous ones and that the system
suffers a continuous bifurcations into new states. In this case, a
degree of coherence has to be preserved between the new and the old
states and this is in agreement with mean-field calculations on the
chaos problem.

It would also be very interesting to understand the chaotic nature of
the low temperature phase in other spin-glass models, random-field
problems and the vortex-glass phase in superconductors. The chaos
approach could reveal as a good starting point to obtain (like happens
in Ising spin glasses) the lower critical dimension for several models
in which there is still much controversy.

From the study of chaoticity in spin glasses we also expect to give some
hint regarding some real dynamical experiments in spin glasses
\cite{Re87,Sa88}. Cycling temperature experiments show that by lowering the
temperature some degree of correlation is preserved between the probed
states\cite{Ha92,Ri94}. Even though experimental spin glasses never reach
equilibrium we think that a correct answer to the statics is relevant to
a qualitative understanding of the effect of perturbations in the
out-off equilibrium relaxations
\cite{Bo92,Cu93}. There are also in course \cite{Ha94} some cycling
magnetic field experiments which we hope will be in agreement with the
main conclusions of this work.

\paragraph{Acknowledgements} 

I gartefully acknowledge very stimulating conversations with A. J. Bray,
S. Franz, D. Huse, M. A. Moore and G. Parisi. This work has been
supported by the European Community (Contract B/SC1*/915198).  I am
grateful to J.  Kurchan, E. Marinari an M. Picco for a careful reading
of the manuscript.
\vfill\eject

\vfill\eject

\begin{center}
FIGURE CAPTION
\end{center}

\vskip1truecm

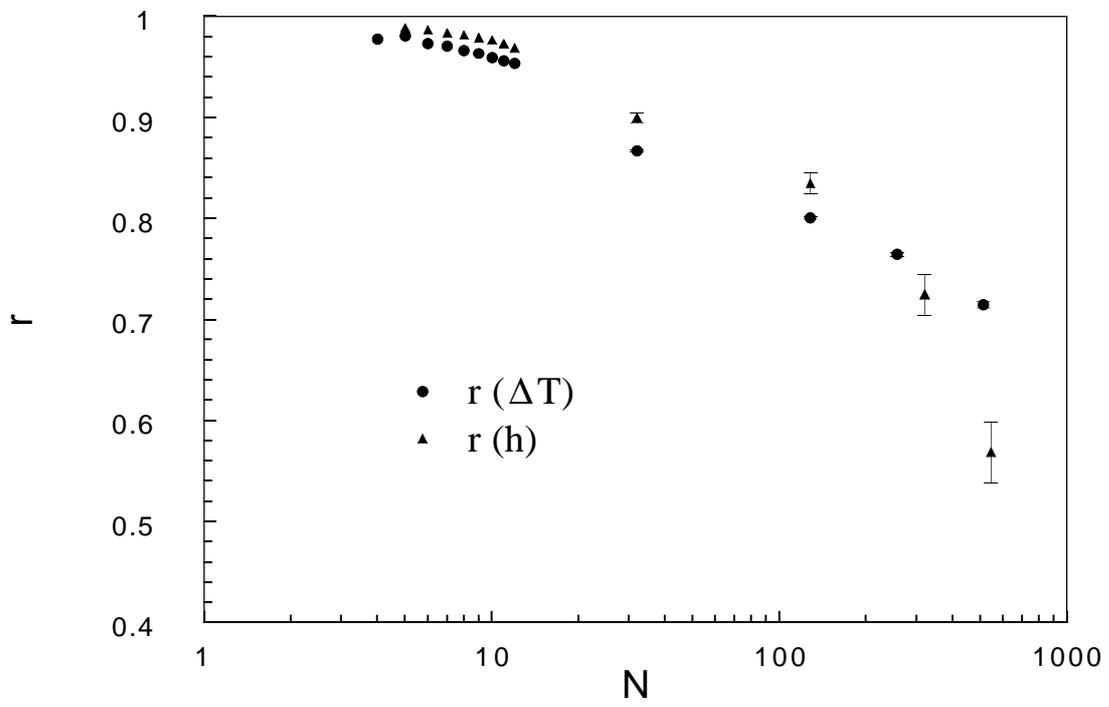
Fig. 1 \quad Chaos parameter $r$ in the SK model for two different
perturbations. In one case the system is at $T=0.6$,$h=0.$ and we apply
a field $h=0.2$, $T$ staying constant. In the other case the system is
at $T=0.5$ and we increase the temperature by $\Delta T=0.4$. Error bars
in the second case are smaller than the size of the symbols. More
details in the text.
\vskip1truecm  

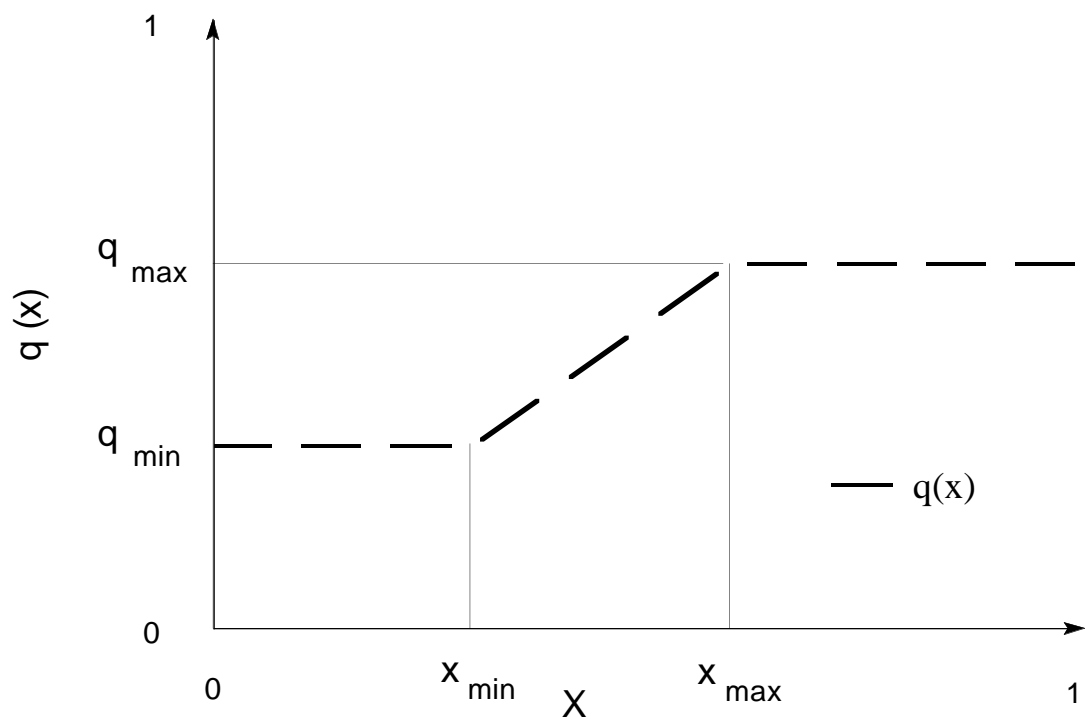
Fig. 2 \quad Mean-field theory order parameter function $q(x)$ in the
spin-glass phase with magnetic field. It is characterized by a minimum
overlap $q_{min}$, a maximum overlap $q_{max}$ and the corresponding
breakpoints $x_{min}$,$x_{max}$.
\vskip1truecm  

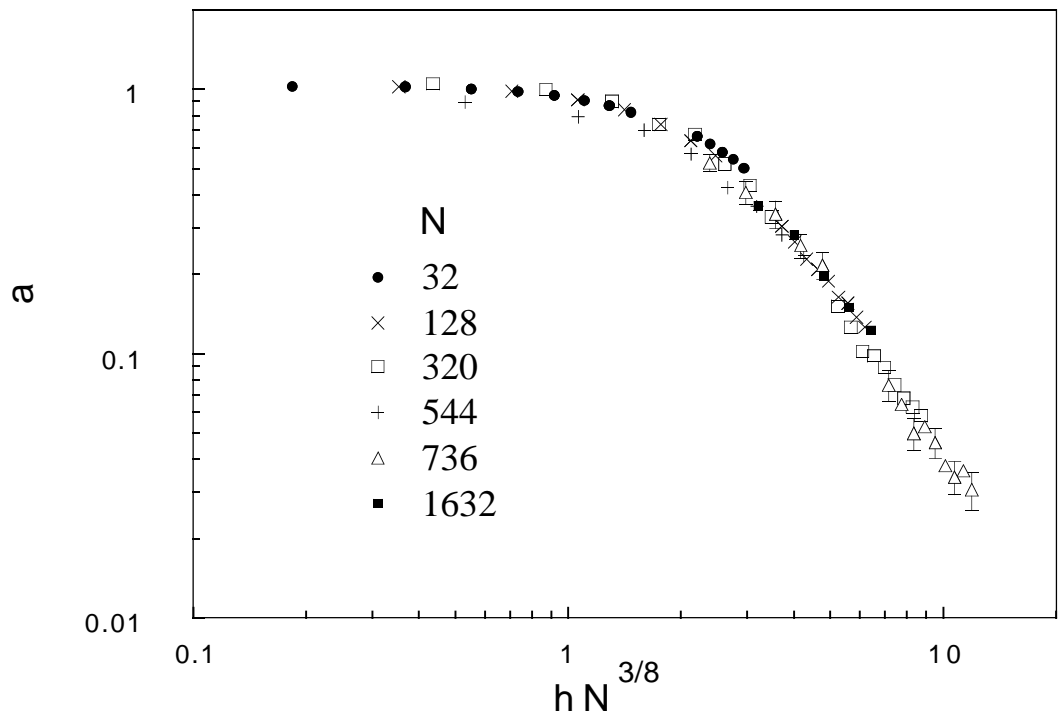
Fig. 3 \quad Chaos with magnetic field in the SK model at $T=0.6$. Field
values range from $h=0.2$ up to $h=1.0$ for the smaller sizes and up to
$h=0.4$ for the largest ones.The number of samples range from $200$ for
$N=32$ down to $40$ for $N=1632$. Typical error bars are of order 5 per
cent in all cases.
\vskip1truecm  

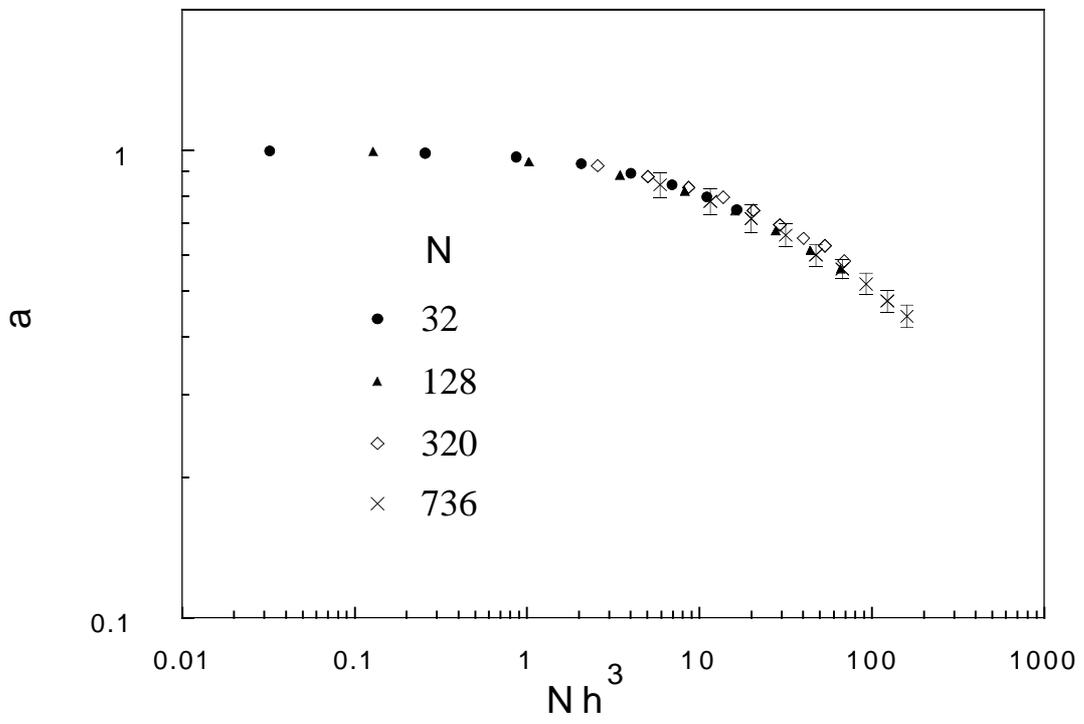
Fig. 4 \quad Chaos with magnetic field in the SK model at $T=T_c=1$.
Field values range from $h=0.1$ up to $h\sim 0.8$.  The number of
samples range from $200$ for $N=32$ down to $50$ for $N=736$.
\vskip1truecm  

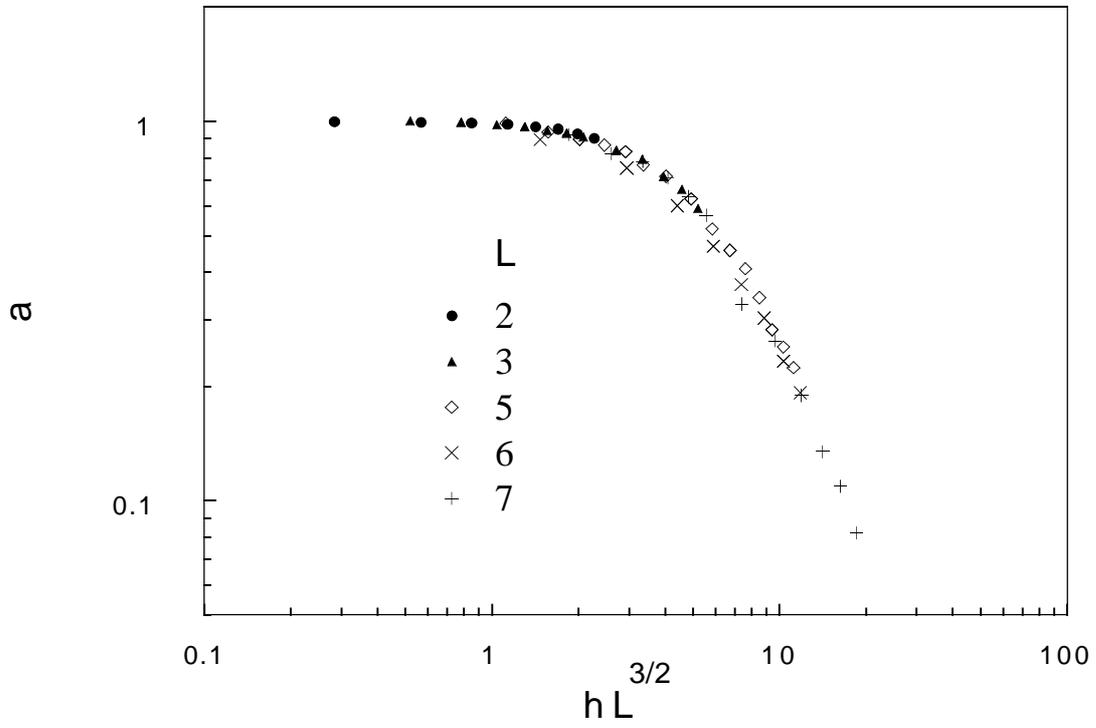
Fig. 5 \quad Chaos with magnetic field in the $4d$ $\pm J$ Ising spin
glass at $T=1.5$. Magnetic field values range from $h=0.1$ up to $h=1$.
The number of samples is approximately $100$ for all lattice sizes.
Typical error bars in this case are of the size of the symbols.
\vskip1truecm  

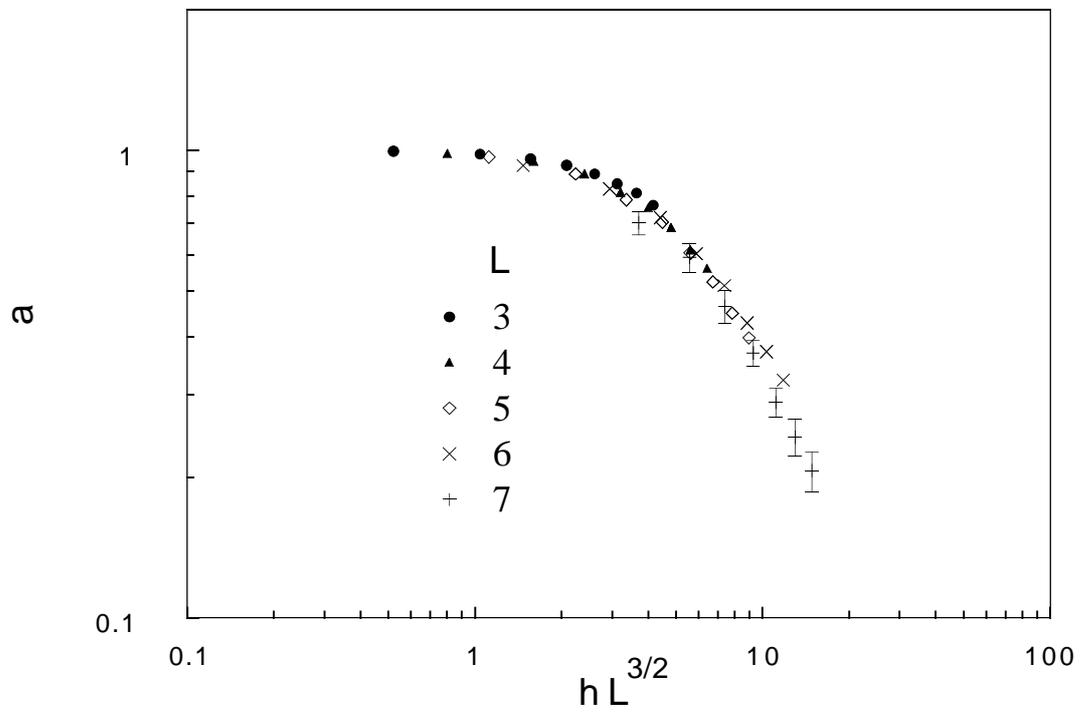
Fig. 6 \quad Chaos with magnetic field in the $3d$ $\pm J$ Ising spin
glass at $T=1.5$. Magnetic field values range from $h=0.1$ up to $h=1.$
The number of samples is approximately $200$ for all lattice sizes
except for $L=7$ in which there are $100$ samples.
Typical error bars are shown in case $L=6$.
\vskip1truecm  

\vfill\eject


\begin{thebibliography}{99}

\bibitem{Sp} M. Mezard, G. Parisi and M. A. Virasoro in ``Spin Glass
Theory and Beyond'' (World Scientific, 1987)

\bibitem{Bi86} K. Binder and A. P. Young,
{\sl Rev. Mod. Phys.} {\bf 58} (1986) 801  

\bibitem{Sk75} D. Sherrington and S. Kirkpatrick,
{\sl Phys. Rev. B} {\bf 17} (1978) 4384

\bibitem{Mi84} W. L. Mc. Millan, 
{\sl J. Phys. C: Solid State Phys.} {\bf 17} (1984) 3179

\bibitem{Br87} A. J. Bray and M. A. Moore, 
{\sl Phys. Rev. Lett.} {\bf 58} (1987) 57

\bibitem{Ko88} G. J. M. Koper and H. J. Hilhorst,
{\sl J. Physique (France)} {\bf 49} (1988) 429  

\bibitem{Fi88} D. S. Fisher and D. A. Huse, 
{\sl Phys. Rev. B} {\bf 38} (1988) 386

\bibitem{Do83} C. De Dominicis and I. Kondor, 
{\sl Phys. Rev. B} {\bf 27} (1983) 606

\bibitem{Do84} C. De Dominicis and I. Kondor, {\sl J.
Physique Lett.} {\bf 45} (1984) L-205

\bibitem{Sl93} G. Parisi, F. Ritort and F. Slanina,
{\sl J. Phys. A: Math. Gen.} {\bf 26} (1993) 247 , 3775

\bibitem{Br84} A. J. Bray and M. A. Moore, 
{\sl J. Phys. C: Solid State Phys.} {\bf 17} (1987) L463

\bibitem{Ka82} S. R. McKay, A. N. Berker, S. Kirkpatrick,
{\sl Phys. Rev. Lett.} {\bf 48} (1982) 767

\bibitem{At78} J. R. de Almeida and D. J. Thouless,
{\sl J. Phys. A: Math. Gen.} {\bf 11} (1978) 983

\bibitem{Me84}M. M\'ezard, G. Parisi, N. Sourlas, G. Toulouse and
M. Virasoro, \it J.Phys. (France) \bf 45 \rm (1984) 843

\bibitem{Ko89} I. Kondor,
{\sl J. Phys. A: Math. Gen.} {\bf 22} (1989) L163

\bibitem{To80} G. Parisi and G. Toulouse,
{\sl J. Phys. Lett.(France)} {\bf 41} (1980) L361

\bibitem{Ni92} M. Nifle and H. J. Hilhorst,
{\sl Phys. Rev. Lett.} {\bf 68} (1992) 2992

\bibitem{Doyo83} C. De Dominicis and A. P. Young,
{\sl J. Phys. A: Math. Gen.} {\bf 16} (1983) 2063

\bibitem{Pa83} G. Parisi,
{\sl Phys. Rev. Lett.} {\bf 50} (1983) 1946

\bibitem{Yo83} A. P. Young,
{\sl Phys. Rev. Lett.} {\bf 51} (1983) 1206

\bibitem{Pa80} G. Parisi,
{\sl J. Phys. A: Math. Gen.} {\bf 13} (1980) 1101,1887

\bibitem{Te88} T. Temesvari, I. Kondor, C. De Dominicis,
{\sl J. Phys. A: Math. Gen.} {\bf 21} (1988) L1145

\bibitem{Pa93} G. Parisi and F. Ritort, 
{\sl J. Phys. A: Math. Gen.} {\bf 26} (1993) 6711 

\bibitem{Ci93} J. C. Ciria,G. Parisi and F. Ritort, 
{\sl J. Phys. A: Math. Gen.} {\bf 26} (1993) 6731 

\bibitem{Do93} C. De Dominicis, I. Kondor and T. Temesvari,
{\sl Int. J. Mod. Phys.} {\bf B7} (1993) 986

\bibitem{Ta92} N. Kawashima, N. Hatano and M. Suzuki, 
{\sl J. Phys. A: Math. Gen.} {\bf 25} (1992) 4985 

\bibitem{Ca94} L. Bernardi and I. A. Campbell,
{\sl received preprint}

\bibitem{Si86} R. R. P. Singh and S. Chakravarty,
{\sl Phys. Rev. Lett.} {\bf 57} (1986) 245

\bibitem{Br80} A. J. Bray and S. A. Roberts,
{\sl J. Phys. C: Solid State Phys.} {\bf 13} (1980) 5405

\bibitem{Gr92} E. R. Grannan and R. E. Hetzel,
{\sl Phys. Rev. Lett.} {\bf 67} (1992) 907

\bibitem{Ru93} J. C. Ciria, G. Parisi, F. Ritort, J. J. Ruiz, 
{\sl J. Physique I (France)} {\bf 3} (1993) 2207 

\bibitem{Ma93} E. Marinari, G. Parisi, F. Ritort 
{\sl preprint cond-mat 9310041}

\bibitem{Bh88} R. N. Bhatt and A. P. Young, 
{\sl Phys. Rev. B} {\bf 37} (1988) 5606

\bibitem{Og85} A. T. Ogielsky,
{\sl Phys. Rev. B} {\bf 32} (1985) 7384

\bibitem{Or93} G. G. Kenning, D. Chu, B. Alavi, J. M. Hamman and R. Orbach,
{\sl J. Appl. Phys.} {\bf 69} (1991) 6240

\bibitem{Pi94} M. Picco and F. Ritort,
{\sl prperint cond-mat 9403077} 

\bibitem{Re87} Ph. Refregier, E. Vincent, J. M. Hamman and M. Ocio,
{\sl J. Physique (France)} {\bf 48} (1987) 1533  

\bibitem{Sa88} L. Sandlund, P. Svendlindh, P. Granberg and P. Norblad,
{\sl J. Appl. Phys.} {\bf 64} (1988) 5616  

\bibitem{Ha92} J. M. Hamman, M. Ledermann, M. Ocio, R. Orbach and E.
Vincent, {\sl Physica A} {\bf 185} (1992) 278

\bibitem{Ri94} H. Rieger, 
{\sl preprint cond-mat 940241}

\bibitem{Bo92} J. P. Bouchaud,
{\sl J. Physique I (France)} {\bf 2} (1992) 1705

\bibitem{Cu93} L. F. Cugliandolo and J. Kurchan,
{\sl Phys. Rev. Lett} {\bf 71} (1993) 173

\bibitem{Ha94} J. M. Hamman, private communication


\end{thebibliography}
\end{document}